\documentclass{PoS}

\title{The Effelsberg Bonn HI Survey EBHIS}

\ShortTitle{The Effelsberg Bonn HI Survey EBHIS}

\author{\speaker{J\"urgen Kerp}, Benjamin Winkel, Peter M.W. Kalberla\\
        Argelander-Institut f\"ur Astronomie, Universit\"at Bonn\\
        E-mail: \email{jkerp@astro.uni-bonn.de},
\email{bwinkel@astro.uni-bonn.de},
\email{pkalberla@astro.uni-bonn.de}}

\abstract{The Effelsberg-Bonn HI Survey (EBHIS) covers the whole sky north of Dec(2000)$\, =\, -5^{\rm \circ}$ on a fully sampled angular grid. Using state-of-the-art FPGA-spectrometers we perform a Milky Way and an extragalactic HI survey in parallel. Moreover, the high dynamic range and short dump time of the HI spectra allow to overcome the vast majority of all radio-frequency-interference (RFI) events. The Milky Way data will be corrected for the stray-radiation bias which warrants a main-beam efficiency of 99\%. Towards the whole survey area we exceed the sensitivity limit of HIPASS, while towards the Sloan-Digital-Sky-Survey (SDSS) area EBHIS offers an order of magnitude higher mass sensitivity. The Milky Way data will be a cornerstone for multi-frequency astrophyics, while the extragalactic part will disclose detailed information on the structure formation of the local universe.}

\FullConference{Panoramic Radio Astronomy: Wide-field 1-2 GHz research on galaxy evolution - PRA2009\\
		 June 02 - 05 2009\\
		 Groningen, the Netherlands}

\begin{document}

\section{The Effelsberg 21-cm multi-beam system}
The seven beam system was primarily optimized to perform Beam-Park experiments. Using the Tracking and Imaging Radar (TIRA) telescope, space debris at an altitude of about 200 km are illuminated. The Effelsberg multi-feed system receives the reflected signal of these space debris. Using the TIRA-Effelsberg combination space debris down to a linear size of 1\,cm can be investigated \cite{Banka}.

\section{EBHIS concept}
Observing time is expensive. Accordingly, we optimized EBHIS to provide valuable data for a broad scientific community. For this purpose, we decided to use the maximum of the available bandwidth and FPGA-spectrometers to obtain high-quality data because of 
\begin{itemize}
\item short integration times of less than a second: this allows us to clean the data for the bulk of the RFI events, because they are variable in time, frequency, and polarization.
\item an order of magnitude more spectral channels than classical auto-correlators, this provides the chance to map narrow spectral lines within a huge bandwidth range. 
\end{itemize}

The maximum EBHIS bandwidth is 100\,MHz. The 14 FPGA spectrometers offer for each polarization 16.384 spectral channels, providing a corresponding velocity resolution of 1.3\,${\rm km\,s^{-1}}$. This is sufficiently narrow to resolve even the coldest CNM-structures belonging to the Milky Way. The 0.5\,sec integration time overcomes, in combination with the 8-bit dynamic range, even the strongest recorded RFI-events. Multiple coverages of the survey area will allow to correct for the contribution of the stray-radiation received by the near and far side-lobes of the single dish. Applying a proper stray-radiation correction allows to reach an accuracy level equivalent to a 99\% main-beam efficency. We are using frequency switch observations with 4\,MHz frequency shift. The covered velocity range is -1000 to +20000$\,{\rm km\,s^{-1}}$.

The extragalactic part of EBHIS is focused on the SLOAN-Digital Sky Survey (SDSS) area. Towards this region we expect to provide a scientifically significant contribution to studies of the local universe. The integration time of ten minutes per beam will provide a mass sensitivity of about $10^7\,{\rm M_\odot}$ at the distance of the Virgo cluster. This will overcome the todays severe statistical limitation of the HI-mass function towards the low-mass end. In comparison to the HIPASS \cite{Barnes}, EBHIS offers a higher sensitivity because of a larger redshift range and a higher velocity resolution across the whole survey area down to Dec = $-5^{\circ}$. Towards the SDSS area we reach an order-of-magnitude higher sensitivity level than HIPASS.  

We observe five by five degrees fields. For the Milky Way survey we use a scanning speed of four degrees per minute. Two coverages are sufficient to reach our final sensitivity limit for the Milky Way survey part accordingly. During the observations the feed orientation is permanently optimized to warrant the full angular sampling. In regular time intervals we observe the standard position S7 for an absolute calibration of our data. In between the noise-diode is used for relative calibration. Test observations disclosed the stability of the whole receiving system lasting for several hours. 

\section{Science goals}
EBHIS will address multiple scientific objectives. Here, we sketch only a few major projects. For the Milky Way part we expect to make significant progress in case of halo studies. Studies of high-velocity and intermediate-velocity clouds will allow to investigate their interaction with the disk--halo interface \cite{KuK09}. Combining EBHIS data with absorption line measurements will disclose the multi-phase/excitation structure of the high-altitude gas and allows to disclose their relation of Lyman-$\alpha$ systems at cosmological distances \cite{benbekhti}. The high spectral and angular resolution in combination with the large survey area will allow for the first time to evaluate the rate at which tiny halo structures appear to populate the lower galactic halo. In the near future, we expect that the soft X-ray all-sky survey of eROSITA allows to detect the bulk of the local baryons by their diffuse soft X-ray emission. EBHIS and GASS (see Kalberla et al. these proceedings) will serve as major resources to evaluate the amount of photoelectric absorption, which is in the order of 80\% to 90\% of the total X-ray flux of the warm-hot-intergalactic medium emission.

For the extragalactic part of EBHIS we will focus on the low-mass end of the HI mass spectrum of the local universe. The high sensitivity combined with our approved RFI-mitigation strategy will allow to derive highly statistical significant information. This will also improve our knowledge on the local baryon budget as well as an evaluation of environmental effects on the HI properties of galaxies in general (see B. Winkel et al., these proceedings).

\section{EBHIS project status}
\begin{figure}
\vspace{3cm}
\centerline{
\includegraphics[width=.9\textwidth]{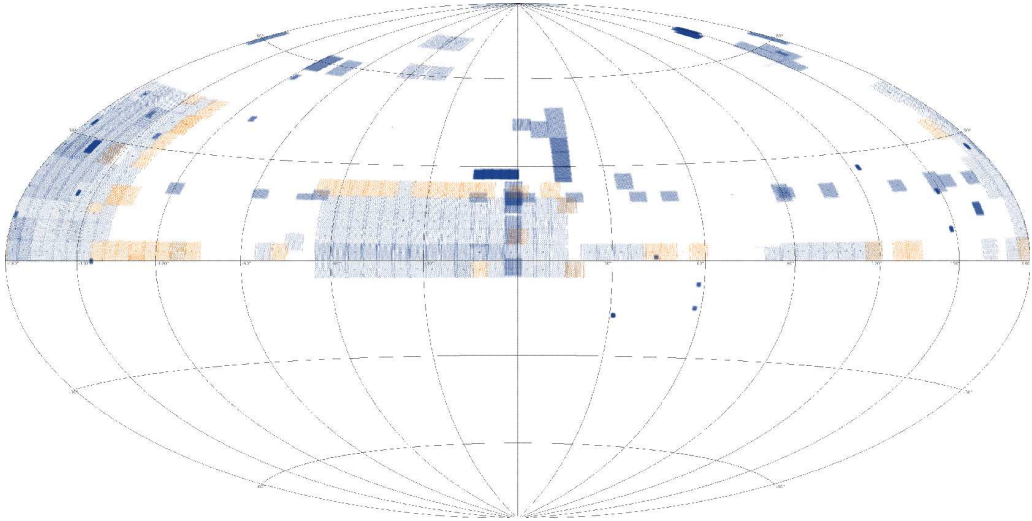}
}
\caption{The plot shows the current (September 2009) coverage of the EBHIS survey. We observe five-by-five degree fields. Up to now fields of interest have been observed until June 2009 where we started the first systematic survey of the northern hemisphere. The map shows a RA-Dec projection with RA(2000)\,$=\,12^{\rm h}$ and Dec(2000)\,$=\,0^{\rm \circ}$ in the center.}
\label{coverage}
\end{figure}

\begin{figure}
\vspace{1cm}
\centerline{
\includegraphics[angle=270,width=.9\textwidth]{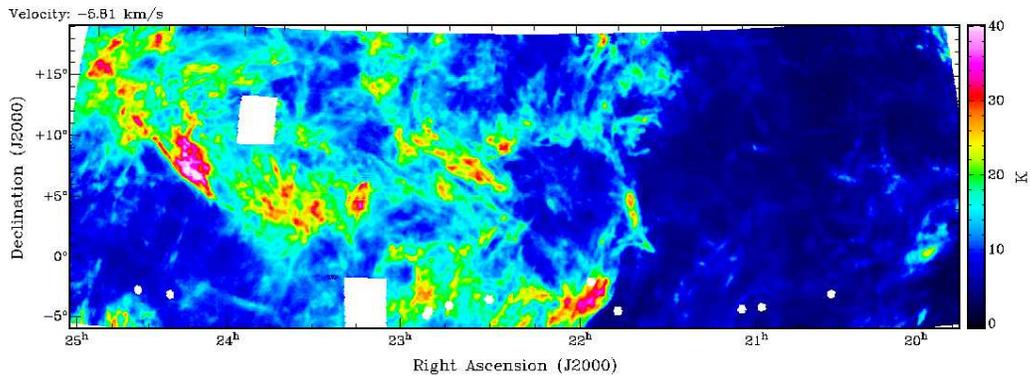}
}
\caption{A channel map of the Milky Way sky displays a wealth of detailed structures. The shown EBHIS Milky Way data are already fully calibrated, corrected for baseline and stray-radiation. RFI events are not removed and cause artifacts at various positions across the map. However, their imprint does not severely degrade the data quality.}
\label{MWcrit}
\end{figure}

During the first observing runs between Christmas 2008 and spring 2009 we spent all observing time to selected areas. These areas were chosen to demonstrate the feasibility of the whole EBHIS project. Accordingly, we focused i.e. on  ``classical'' deep fields and high priority objects. 
Starting in June 2009 we changed our observing strategy to perform the first coverage of the northern Milky Way sky. Figure\,\ref{coverage} shows the present coverage (September 2009) of EBHIS. Figure\,\ref{MWcrit} shows the Milky Way data quality after a quick-and-dirty data reduction, in particular no RFI-mitigation was performed.

\section{EBHIS performance}
\begin{figure}
\vspace{2cm}
\centerline{
\includegraphics[angle=0,width=.6\textwidth]{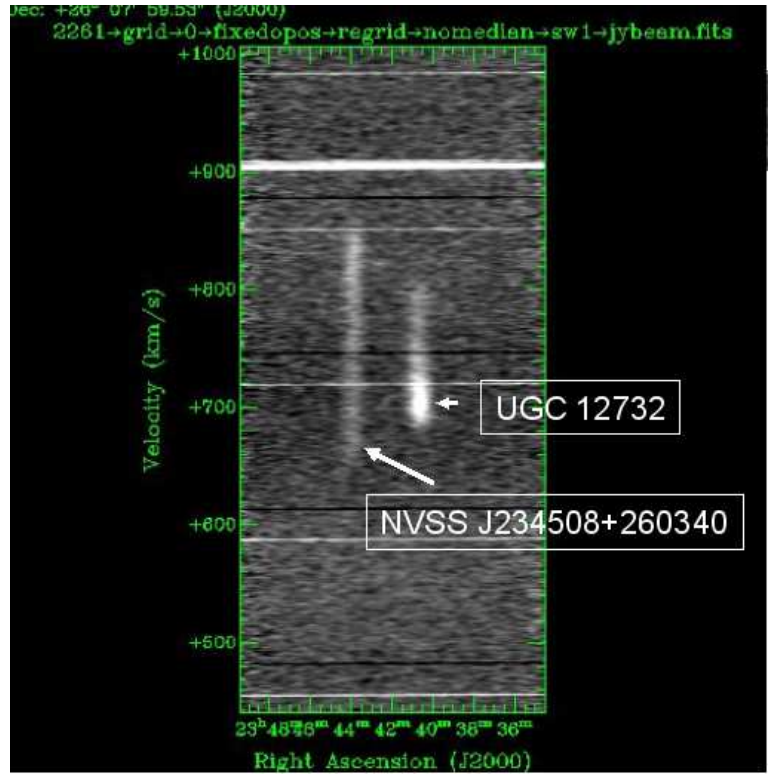}
}
\caption{Position-velocity (PV) map towards the galaxy UGC\,12732. The EBHIS data are fully calibrated and a baseline has been subtracted. The HI spectra show clearly the HI emission of the galaxy. The bright stripes across the PV-map show RFI events, the corresponding dark lines are their frequency-switch counterparts. The galaxy NVSS\,J234508+260340 next to UGC\,12732 is also significantly detected but not part of the ALFALFA catalog of galaxies towards this area of interest.}
\label{galaxies}
\end{figure}

Figure\,\ref{galaxies} shows a position-velocity slice of an area towards the M81/M82-group. At a radial velocity of about 750\,${\rm km\,s^{-1}}$ two galaxies have been detected. UGC\,12732 is detected by the ALFALFA survey \cite{ALFALFA}, while NVSS\,J234508+260340 is not part of the ALFALFA catalog. Obviously, EBHIS detected both galaxies at a high signal-to-noise level. The shown data comprise a net-observing time of four minutes per beam only.

\begin{figure}
\vspace{4cm}
\centerline{
\includegraphics[angle=0,width=.4\textwidth]{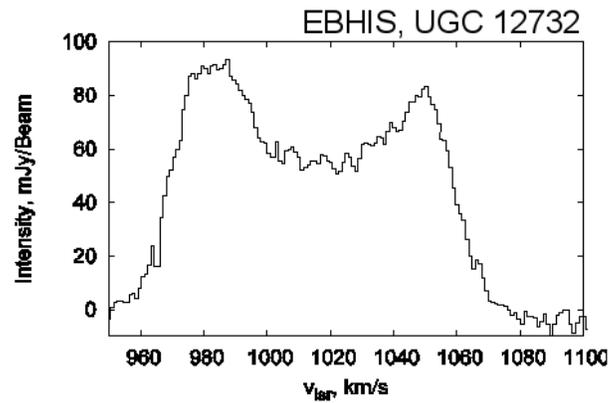}
}
\caption{EBHIS HI spectrum of UGC\,12732. The spectrum shows the quality of the HI data obtained. The integration time for this spectrum was 4\,minutes. Applying our RFI-mitigation and increasing the net observing time towards the SDSS area will provide a unique data base for studies of the local universe.}
\label{UGCHIspectrum}
\end{figure}

Figure\,\ref{UGCHIspectrum} emphasizes the high velocity resolution of EBHIS thanks to the FGPA-spectrometers and their high number of spectral channels. The channel number is important with respect to the RFI-mitigation because most RFI events are narrow in frequency accordingly only a small portion of the galaxy rotation curve is affected. A comparison between the derived fluxes between ALFALFA and EBHIS show a significant discrepancy. This is an open issue which we have to solve in the near future.  

The EBHIS Milky Way survey will be finished within two years of observing time, assuming 1000 hours of time per year. The extragalactic survey will need five years of observing time to reach its ultimate sensitivity limit.

The EBHIS data will be distributed via an Internet interface comparable to the LAB survey \cite{LAB}. In addition, we plan to provide data cubes for scientific use.


\acknowledgments The authors thank the Deutsche Forschungsgemeinschaft (DFG) for financial support under the research grant KE757/7-1. Based on observations with the 100-m telescope of the MPIfR (Max-Planck-Institut f\"ur Radioastronomie) at Effelsberg.

\end{document}